\documentclass[twocolumn,prl,showpacs]{revtex4}
\usepackage{amssymb, amsmath, graphicx}
\def\be{\begin{equation}}
\def\ee{\end{equation}}
\begin{document}
\title{Time-Average Based on Scaling Law in Anomalous Diffusions
}
\author{Hyun-Joo \surname{Kim}}\email{hjkim21@knue.ac.kr}
\affiliation{
Department of Physics Education, Korea National University of Education,
Chungbuk 363-791, Korea\\
}

\received{\today}
%\keywords{}
\begin{abstract} 
To solve the obscureness in measurement brought about from the weak ergodicity breaking appeared 
in anomalous diffusions we have suggested the time-averaged mean squared displacement (MSD) 
$\overline{\delta^2 (\tau)}_\tau$ with a integral interval depending  linearly on the lag time
$\tau$. For the continuous time random walk describing a subdiffusive behavior, we have found that 
$\overline{\delta^2 (\tau)}_\tau \sim \tau^\gamma$ like that of the ensemble-averaged MSD, 
which makes it be possible to measure the proper exponent values through time-average 
in experiments like a single molecule tracking. Also we have found that it is originated from 
the scaling nature of the MSD at a aging time in anomalous diffusion and confirmed them  
through numerical results of the other microscopic non-Markovian model showing subdiffusions
and superdiffusions with the origin of memory enhancement.

\end{abstract}


\maketitle

Ergodic hypothesis has become a keystone by allowing us to replace a time average of a physical 
observable by a ensemble average in statistical physics.  
However, it is rather reasonable that the ergodic hypothesis is not valid for nonstationary 
processes and thus recently in the experiment of the fluorescence intermittency of a nanocrystal which is
governed by anomalous diffusions with Le{\'v}y statistics, it was shown that ensemble-averaged 
properties are different from the time-averaged and suggested the statistical aging
and nonergodicity \cite{PRL2003}. Since then the weak ergodicity breaking in anomalous diffusions
has been found through much more experiments and theoretical models \cite{PRL2005,PRL2006,PRL2013-1,PRL2013-2,PRL2013-3,
PRE2012,PRE2013,PRE2014,PRE2014-1,PNAS2011,PRL2011,PRE2009,PRL2008,PRL2008soko,PRL2007,NJP}.

An anomalous diffusive process is characterized by the nonlinear behavior of the ensemble-averaged 
MSD which grows as the power-law, $\langle x^2 (t) \rangle \sim t^{\gamma}$, which
is comparable with the linear behavior of the MSD in normal diffusion \cite{ad-soko,ctrw,firststep}.
The exponent $\gamma$ classifies superdiffusion ($ \gamma > 1 $) in which the past and future 
random steps are positively correlated and thus persistence is exhibited, and subdiffusion 
($ 0 < \gamma < 1 $) which behaves in the opposite way, showing antipersistence.
The ergodic property in an anomalous diffusion is dealt with on the basis of the time-averaged MSD 
calculated as 
\be
\overline {\delta^2 (\tau)}_T = \frac{1}{ T- \tau} \int_0 ^{T-\tau} [ x(t+\tau) - x(t) ]^2 dt,
\ee
where $\tau$ is the lag time.
For normal diffusion and the long time limit $\overline{\delta^2 (\tau)}_T$ depends linearly on
the lag time $\tau$, i.e., $\overline{\delta^2 (\tau)}_T = 2D\tau$ where $D$ is the diffusion
constant, which is the same as the behavior of ensemble averaged MSD $\langle x^2 (t) \rangle
= 2Dt$, providing an ergodic behavior. 
While for the continuous time random walk (CTRW) with waiting time probability density function lacking
the mean waiting time $\psi(t) \sim {\gamma A}t^{-(1+\gamma)}/{\Gamma (1+\gamma)}$
which describes well subdiffusions shown in various experiments \cite{ad-soko,ctrw}, 
it was found that the ensemble and time averaged MSD behaves as
$\langle \overline {\delta^2 (\tau)}_T \rangle \sim {2D_\gamma \tau}/{\Gamma (1+\gamma) T^{1-\gamma}}$ 
with $0 < \gamma < 1$ in the limit $\tau \ll T$ which is different from the behavior found by an ensemble average, 
$\langle x^2 (t) \rangle = 2D_\gamma t^\gamma / \Gamma(1+\gamma)$
indicating the weak ergodicity breaking due to the aging effect of a nonstationary process \cite{PRL2008,PRL2008soko}.

This weak ergodicity breaking makes difficulties in interpreting data averaged over time for a few trajectories 
such as in a single molecule measurement, in other words, what appears as normal diffusion for time average
may actually be a hidden subdiffusion and the exact value of exponent $\gamma$ can not be known
and thus it gives fundamental obstacles analyzing anomalous diffusions in real experiments.
Therefore finding a method of time average to disclose hidden natures of a diffusion is meaningful and helpful
for measurement in real experiments.
In this study, we provide a time average with different interval of integration for each lag time $\tau$
to solve the obstacles from nonergodic property of anomalous diffusive processes and show
that it gives the same anomalous exponent $\gamma$ as the ensemble-averaged by the scaling property
of the MSD at a aging time $t$. It is considered through the CTRW and the microscopic non-Markovian model describing 
both superdiffusions and subdiffusions by memory enhancement and the heterogeneity of the provided
time-averaged observable for different realizations is considered as well.

We define a time average of MSD with upper limit of integration dependent on the lag time 
as follows,
\be
\overline{ {\delta^2 (\tau)} }_\tau \equiv \frac{1}{ a \tau} \int_0 ^{a\tau} [ x(t+\tau) - x(t) ]^2 dt,
\label{atau}
\ee
where $a$ is a positive constant. To the end of comparison with 
$\langle \overline{ {\delta^2 (\tau)} }_T \rangle$ we first consider the ensemble average of Eq. (\ref{atau}), 
$\langle \overline{ {\delta^2 (\tau)} }_\tau \rangle$.
For the unbiased CTRW model with $\psi(t) \sim {\gamma A}t^{-(1+\gamma)}/{\Gamma (1+\gamma)}$, 
the MSD at a aging time $t$ defined by 
$\langle {\delta^2 (t,\tau)} \rangle = \langle [ x(t+\tau)-x(t)]^2 \rangle$ is given by
$\langle {\delta^2 (t,\tau)} \rangle = \langle l^2 \rangle [ \langle  n(t+\tau) \rangle
- \langle n(t) \rangle]$, where $\langle l^2 \rangle$is the average of jump lengths in the interval 
$(t, t+\tau)$ and $\langle n(t) \rangle $, the average number of jumps during time $t$ 
scales as $\langle n(t) \rangle \sim t^\gamma /A \Gamma (1+\gamma)$ \cite{PRL2008,PRL2008soko}. 
Then we find 
\be
\langle \overline{ {\delta^2 (\tau)} }_\tau \rangle = \frac{2D_\gamma}{\Gamma(1+\gamma)}
\frac{(a+1)^{\gamma+1} - (a^{\gamma+1} + 1)}{a(\gamma+1)} \tau^\gamma \sim \tau^\gamma,
\label{ctrwdelta}
\ee
where $D_{\gamma} = {\langle l^2 \rangle}/{2 A}$.
$\langle \overline{ {\delta^2 (\tau)} }_\tau \rangle$ follows the power-law behavior like
the  ensemble-averaged MSD with the same exponent, i.e., the ensemble average can be replaced by the time average
in obtaining the exponent $\gamma$. In the limit of $a \gg 1$  we obtain
\be
\langle \overline{ {\delta^2 (\tau)} }_\tau \rangle \sim  \frac{2D_\gamma}{\Gamma(1+\gamma)}\frac{ \tau^\gamma}{a^{1-\gamma}}.
\ee
Since the large $a$ corresponds to the long time limit it is plausible that 
$\langle \overline{ {\delta^2 (\tau)} }_\tau \rangle \sim a^{\gamma-1}$
compared to $\langle \overline{ {\delta^2 (\tau)} }_T \rangle \sim T^{\gamma-1}$ indicating that  the longer 
measurement time goes on, the smaller diffusion constant is in the subdiffusion by the CTRW. 
However, taking not fixed measurement time 
but one dependent linearly on the lag time prevents the distortion of the power law behavior of MSD generated by
averaging over time unlike the behavior of $\langle \overline{ {\delta^2 (\tau)} }_T \rangle \sim \tau$.
It may be due to the time scale-free nature in anomalous diffusion.
In a nonstationary process $\langle \overline{ {\delta^2 (\tau)} }_\tau \rangle$ is usually 
different from $\langle { {\delta^2 (0, \tau)} } \rangle$ that is $\langle x^2 (t) \rangle$ for a
process with $\langle x(0) \rangle =0$ where $t$ becomes the lag time from $t=0$, which is due to the aging effect of 
$\langle {\delta^2 (t,\tau)} \rangle$, in other words, when the measurement starts not at time $t=0$
but at a later time $t$ the process depends on the aging time $t$. However, if $\langle {\delta^2 (t, \tau)} \rangle$ 
appears in the form of a scaling function $f(t/\tau)$ like 
\be
\langle {\delta^2 (t,\tau)} \rangle \sim \tau^\gamma f(\frac{t}{\tau}),
\ee
then
\be
\langle \overline{ {\delta^2 (\tau)} }_\tau \rangle \sim \frac{\tau^\gamma}{a } \int_0 ^{a} f(x) dx \sim \tau^\gamma,
\ee
where $ x = t/\tau$. For the CTRW $\langle {\delta^2 (t, \tau)} \rangle \sim (t+\tau)^\gamma - t^\gamma \sim
\tau^\gamma f({t}/{\tau})$ with $f(t/\tau) = (t/\tau +1)^\gamma - (t/\tau)^\gamma$  
so that the power-law behavior of Eq. (\ref{ctrwdelta}) is accomplished.
Thus even for a nonstationary process if a observable has a scaling property and there is a proper scaling function of $t$ and $\tau$, 
the power-law behavior of a observable remains for the time average.
%%%%%%%%%%%%%%%%%%%%%%%%%%%%%%%%%%%%%%%%%%%%%%%%%%%%%%%%%%%%%%%%%%%%%%%%%%%%%%%%%%%%%%   
\begin{figure}[ht]
 \includegraphics[width=9cm]{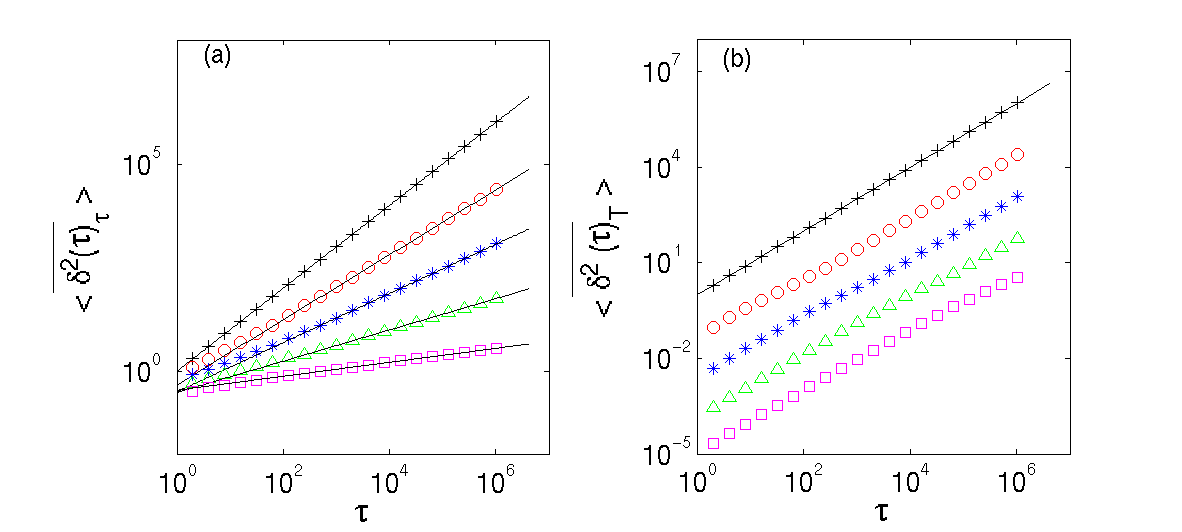}
 \caption{(a) $\langle \overline{\delta^2 (\tau)}_\tau \rangle$ versus $\tau$ with $a=10$ for various
 $\alpha =0, 0.2, 0.4, 0.6,$ and $0.8$ from the top to the bottom. The solid lines are
 the fitting lines for the range $2^{10} \leqslant \tau \leqslant 2^{20}$ with the exponent
 $\gamma \approx 1, 0.78, 0.59, 0.37,$ and $0.18$, respectively.
 (b) $\langle \overline{\delta^2 (\tau)}_T \rangle$ versus $\tau$ with $T=10 \times 2^{20}$ for various
 $\alpha =0, 0.2, 0.4, 0.6,$ and $0.8$ from the top to the bottom.
 The slope of solid line is one, which represents $\langle \overline{\delta^2 (\tau)}_T \rangle
 $ depends linearly on $\tau$ irregardless of the different values of $\alpha$ unlike the results
 of the ensemble averages.  
 }
 \label{sub_dave}
\end{figure}
%%%%%%%%%%%%%%%%%%%%%%%%%%%%%%%%%%%%%%%%%%%%%%%%%%%%%%%%%%%%%%%%%%%%%%%%%%%%%%%%%%%%%%  

Meanwhile in order to consider the heterogeneity of different realizations found in many experiments 
\cite{PRL2003,PRL2006,PNAS2011}
we consider the parameter of the relative dispersion of a observable as a measure of the heterogeneity
\cite{PRL2008,PRE2014}. If it vanishes in the long time limit the system 
is ergodic for stationary processes \cite{st-ergo} and if a process is not ergodic, it has a finite value
even in the long time limit, by which it is called the ergodicity breaking (EB) parameter. 
However, even in the case the parameter vanishes
nonstationarity of a system can induce the ergodicity breaking \cite{PRE2014-1}.
Thus we take the EB parameter $J_\tau$ of $\langle \overline{\delta^2 (\tau)}_\tau \rangle$ as follows,
\be
J_\tau (\tau) \equiv \frac{\langle \overline{\delta^2 (\tau)}^2 _\tau \rangle - 
\langle \overline{\delta^2 (\tau)}_\tau \rangle ^2}{\langle \overline{\delta^2 (\tau)}_\tau \rangle ^2}.
\ee
For the CTRW $\overline{\delta^2 (\tau)}_\tau = 2AD_\gamma n(a \tau)/a$ in the long time limit of $a \gg 1$
\cite{PRE2013} and then we find
\be
J_\tau (\tau) = \frac{2 \Gamma^2 (1+\gamma)}{\Gamma(2\gamma +1)} - 1.
\ee 
It means that even for enough large value of $a$ the process is not homogeneous, 
that is, although $\overline{\delta^2 (\tau)}_\tau$ can give the proper values of anomalous exponent
the heterogeneity remains due to the nonstationarity of the CTRW. 

To make further progress of the above argument for the other anomalous diffusion comprising both superdiffusion 
and subdiffusion we consider a microscopic non-Markovian model \cite{me} 
which well describes anomalous diffusions by the another origin of memory enhancement. 
In the model a walker starts at origin and moves to the right or left with equal probability at time $t=1$
and increment of a jump $\sigma(t)$ at time $t > 1$ is given by
\be
\sigma_{t} = \left \{
\begin{array}{ll}
-\sigma_{t-1},  & \text{with probability $1-1/t^{\alpha}$}\\
1 \; \text{or} \; -1,  & \text{with probability $1/t^{\alpha}$}
\end{array} \right.
\label{stm}
\ee
Over time, the probability of taking the opposite direction of the latest step increases with time
and the larger value of parameter $\alpha$ is, the much faster the probability grows. 
That is, the anti-persistence with the previous step is enhanced with time of which degree is 
controlled by the parameter $\alpha$ and it was found that $\gamma = 1-\alpha$. 
When $\alpha =0$ it reduced to the original random walk
and the subdiffusive behaviors are induced with the strength controlled by the parameter $\alpha$.

Figure \ref{sub_dave} (a) shows the plot of $\langle \overline{\delta^2 (\tau)}_\tau \rangle$ 
versus $\tau$ with $a=10$ for various $\alpha =0, 0.2, 0.4, 0.6,$ and $0.8$. Simulation results were 
obtained by 1000 independent realizations for all below appeared data. 
The fitted solid lines represent that $\langle \overline{\delta^2 (\tau)}_\tau \rangle \sim \tau^{\gamma}$
as shown in the results obtained by ensemble averages \cite{me}. 
While $\langle \overline{\delta^2 (\tau)}_T \rangle \sim \tau$ irrespective of the values of $\alpha$
(Fig. \ref{sub_dave} (b)), which indicates the ergodicity breaking due to the nonstationarity
like the case of CTRW.
Thus for the non-Markovian walk model $\langle \overline{\delta^2 (\tau)}_T \rangle$ 
can not also provide proper information for a anomalous diffusion, while 
$\langle \overline{\delta^2 (\tau)}_\tau \rangle$ gives the proper values of
$\gamma$ being able to describe anomalous subdiffusions. It means that
$\langle \overline{\delta^2 (\tau)}_\tau \rangle$ can be served as a measurable quantity to confirm
anomalous diffusions through time average in experiments with a few trajectory such as the single molecule tracking.   

\begin{figure}[ht]
 \includegraphics[width=8cm]{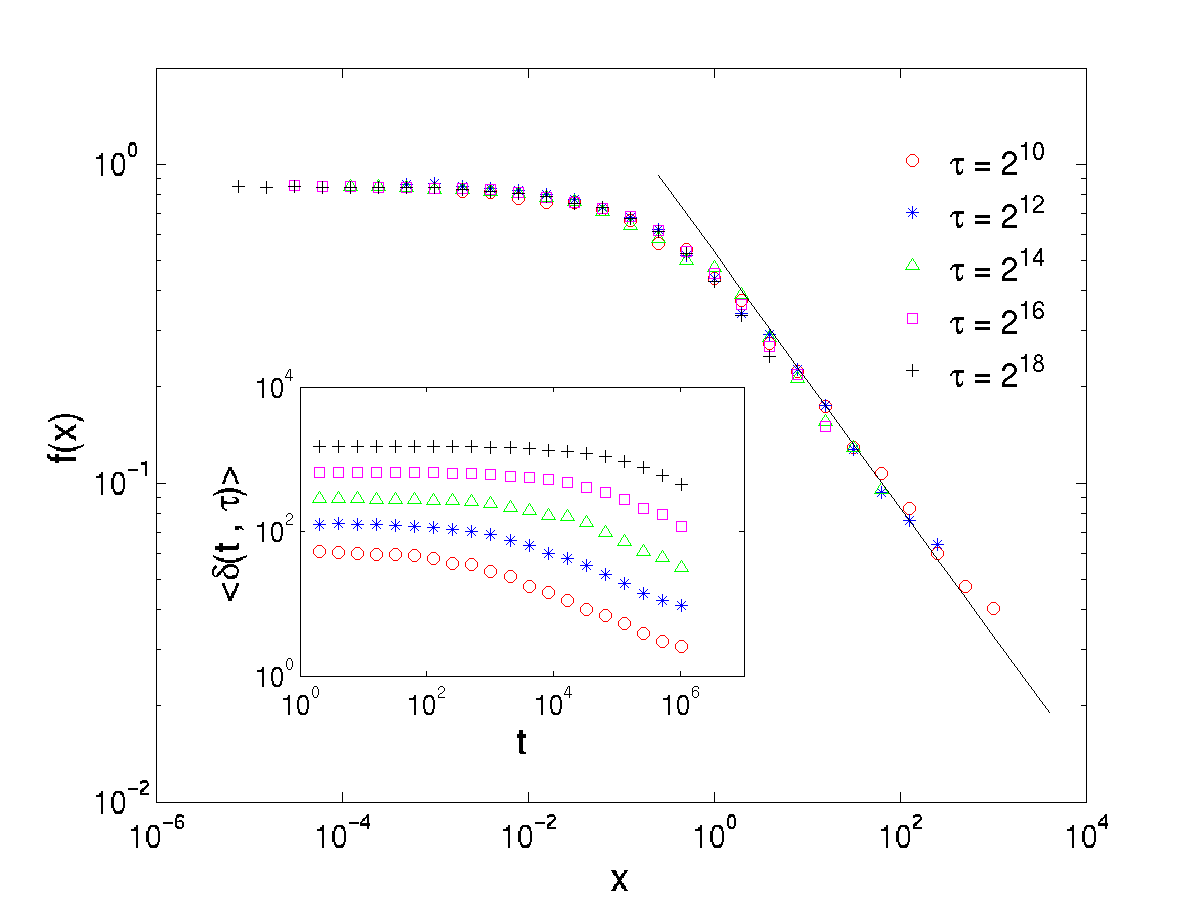}
 \caption{Scaling functions $f(x)$ versus $x$ for the different fixed values of $\tau$
 with $\alpha = 0.6$,
 which fall well into a single curve. The slope of solid line is $-0.4$, which represents 
 that $f(x) \sim x{^-\beta}$ with $\beta = 1-\gamma$. 
 Inset shows the plot of $\langle \delta^2 (t, \tau)\rangle$ versus time $t$
 for the various values of fixed $\tau$.
 }
 \label{sub_fx}
\end{figure}

The scaling behavior of $\langle {\delta^2 (t, \tau)} \rangle$ of the model is shown in the Fig. \ref{sub_fx}.
$\langle {\delta^2 (t, \tau)} \rangle$ for different fixed lag times collapse
very well in a single curve, indicating $\langle {\delta^2 (t, \tau)} \rangle \sim \tau^\gamma f(t/\tau)$,
where 
\be
f(x) \sim  \left \{
\begin{array}{ll}
\text{const.},  & \quad \text{for $x \ll 1$}\\
x^{-\beta},  & \quad \text{for $x \gg 1,$}
\end{array} \right.
\label{fx}
\ee
with $\beta = 1-\gamma$. $\langle {\delta^2 (t, \tau)} \rangle \sim \tau^\gamma$ for $t \ll \tau$, which means
that when the aging time is much shorter than the observation time interval the aging effect is ignored
while for $t \gg \tau$, $\langle {\delta^2 (t, \tau)} \rangle \sim \tau / t^{1-\gamma}$, i.e.
when the aging time is much longer than the observation time interval the process is reduced in a normal
diffusion with the diffusion coefficient proportional to the aging time as $t^{\gamma-1}$ 
(the inset of Fig. \ref{sub_fx}).
It results in the normal behavior of $\langle \overline{\delta^2 (\tau)}_T \rangle$
in the long time limit. These results are the same as those of the CTRW showing the subdiffusive
behavior \cite{firststep}. Thus the subdiffusive property can be extracted by $\langle \overline{\delta^2 (\tau)}_\tau \rangle$
from the underlying scaling nature of  $\langle {\delta^2 (t, \tau)} \rangle$.    

%%%%%%%%%%%%%%%%%%%%%%%%%%%%%%%%%%%%%%%%%%%%%%%%%%%%%%%%%%%%%%%%%%%%%%%%%%%%%%%%%%%%%%   
%%%%%%%%%%%%%%%%%%%%%%%%%%%%%%%%%%%%%%%%%%%%%%%%%%%%%%%%%%%%%%%%%%%%%%%%%%%%%%%%%%%%%%
%%%%%%%%%%%%%%%%%%%%%%%%%%%%%%%%%%%%%%%%%%%%%%%%%%%%%%%%%%%%%%%%%%%%%%%%%%%%%%%%%%%%%%     
\begin{figure}[ht]
 \includegraphics[width=9.5cm]{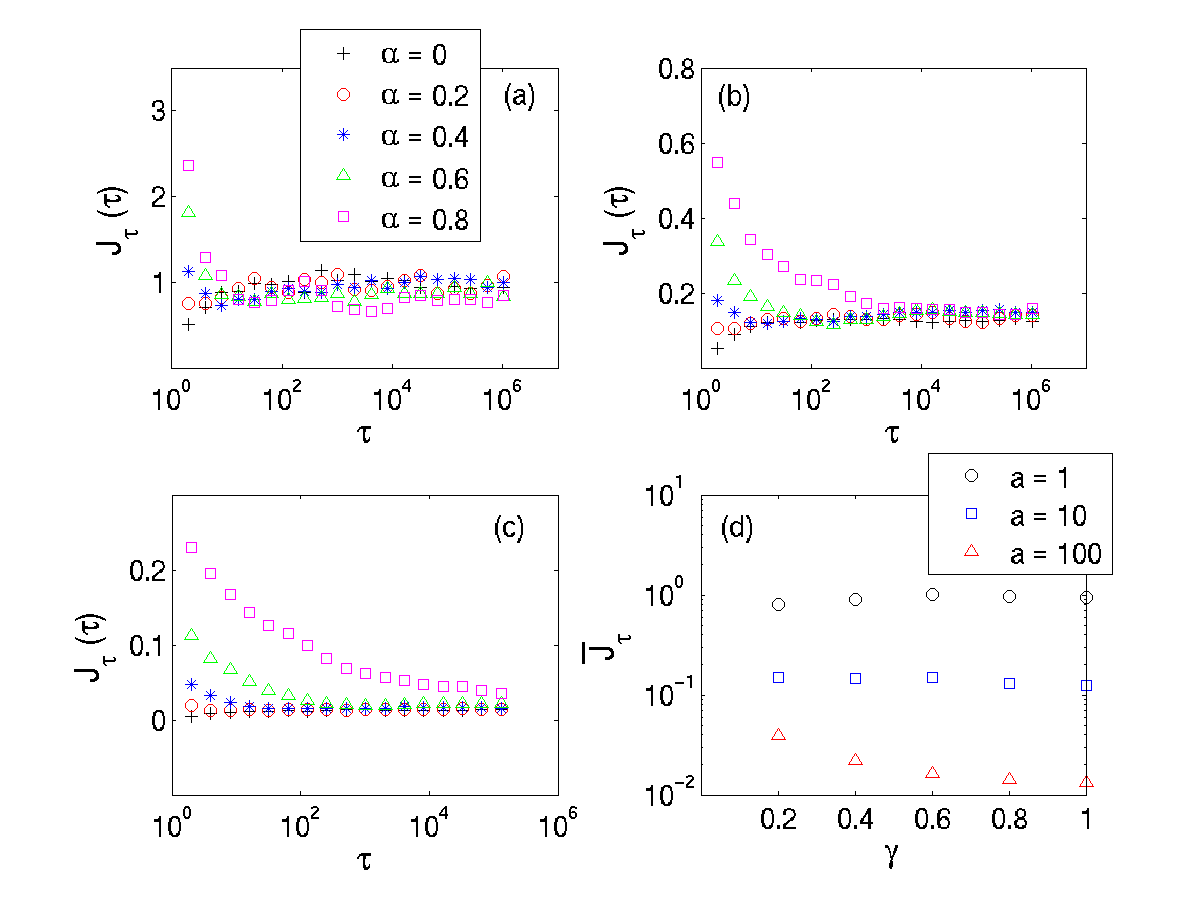}
 \caption{$J_\tau (\tau)$ as a function of $\tau$ for 
 $\langle \overline{\delta^2 (\tau)}_\tau \rangle$ with various $\alpha$ and (a) $a=1$,
 (b) $a=10$, and (c) $a=100$.
 (d) $\bar J_\tau$ averaged for the range $2^{15} \leqslant \tau \leqslant 2^{20}$ with 
 various $a$ for the exponent $\gamma$.
 }
 \label{sub_hete}
\end{figure}
%%%%%%%%%%%%%%%%%%%%%%%%%%%%%%%%%%%%%%%%%%%%%%%%%%%%%%%%%%%%%%%%%%%%%%%%%%%%%%%%%%%%%%  
Figure \ref{sub_hete} shows $J_\tau (\tau)$ as a function of $\tau$ with (a) $a=1$, (b) $a=10$, and 
(c) $a=100$ for different values of $\alpha$. For large $\tau$, 
$J$ does not depend on $\tau$ and such behavior is more delayed when $\alpha$ is larger.
The larger $\alpha$ is the stronger strength of ant-persistence is, which results in the more
heterogeneity in the range of small $\tau$. 
Also as shown in the Fig. \ref{sub_hete} (d) the values of $J_\tau$ averaged over the large $\tau$ 
decrease when $a$ increases, that is, as the interval
of time average increases the heterogeneity decreases and for all $a$ the values for $\gamma < 1$
are close to that of $\gamma=1$ indicating the normal behavior, which means that 
the heterogeneity does not result from the nontstationarity of subdiffusive behavior but
the finite time effect and thus in the long time limit ($a \rightarrow \infty$) it is expected
that the heterogeneity could be disappeared. 
Thus $\overline{\delta^2 (\tau)}_\tau$ shows the homogeneity for different realizations 
in the long time limit unlike that of the CTRW and thus the heterogeneity of $\overline{\delta^2 (\tau)}_\tau$
may depend on models.

%%%%%%%%%%%%%%%%%%%%%%%%%%%%%%%%%%%%%%%%%%%%%%%%%%%%%%%%%%%%%%%%%%%%%%%%%%%%%%%%%%%%%%   
\begin{figure}[ht]
 \includegraphics[width=9cm]{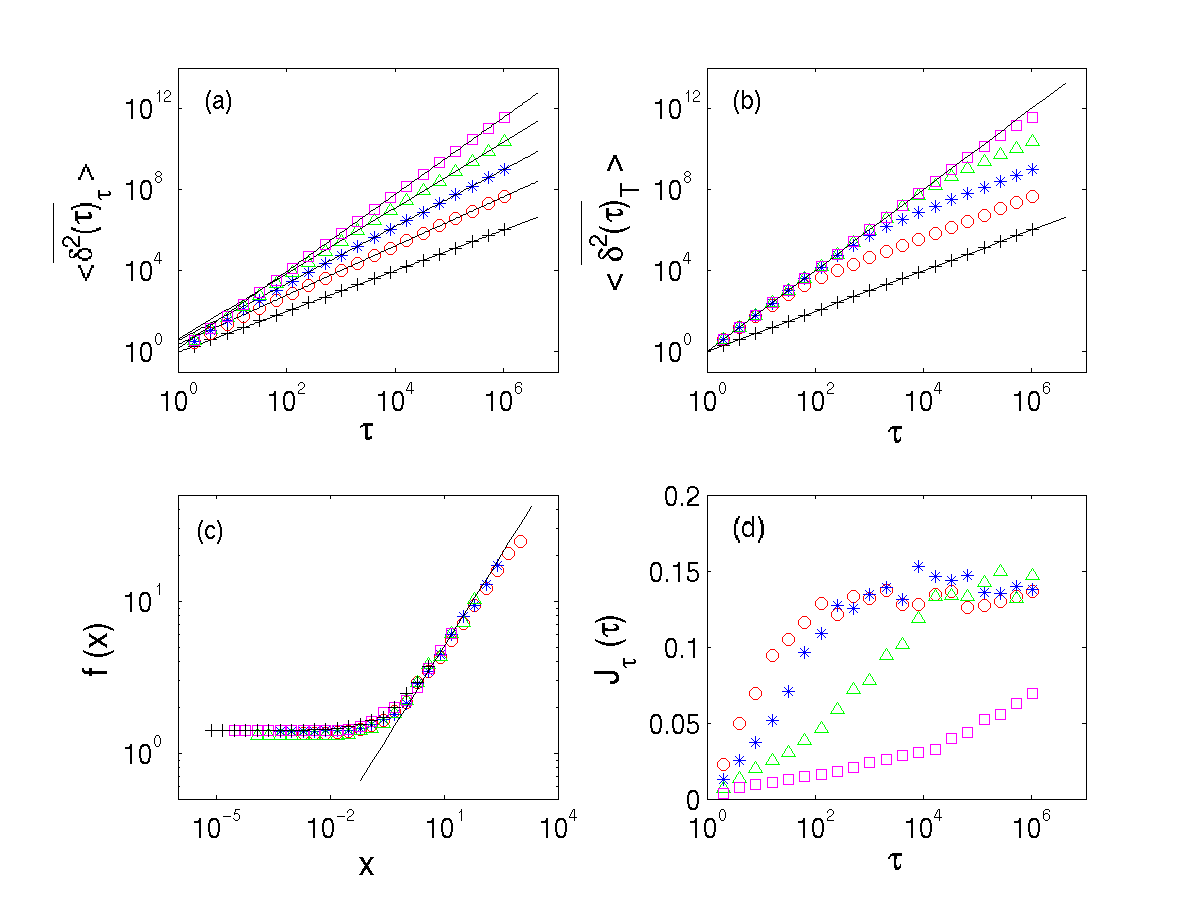}
 \caption{Data in this figure were obtained for the model for a superdiffusive behavior.
 (a) $\langle \overline{\delta^2 (\tau)}_\tau \rangle$ versus $\tau$ with $a=10$ for various
 $\alpha =0, 0.2, 0.4, 0.6,$ and $0.8$ from the bottom to the top. The solid lines represent
 that $\gamma \approx 1, 1.21, 1.41, 1.62,$ and $1.90$, respectively.
 (b) $\langle \overline{\delta^2 (\tau)}_T \rangle$ versus $\tau$ with $T=10 \times 2^{20}$ for various
 $\alpha =0, 0.2, 0.4, 0.6,$ and $0.8$ from the bottom to the top.
 The slope of solid line is 1 and the slope of dotted line is 2, which represents 
 $\langle \overline{\delta^2 (\tau)}_T \rangle$ crossovers from the ballistic motion to the normal 
 diffusive behavior.
 (c) Scaling function $f(x)$ versus $x$ for the different fixed values of $\tau$ with $\alpha = 0.4$.
 They fall well into a single curve satisfying the Eq. (\ref{fx}) with $\beta = 1-\gamma = -0.4$.
(d) EB parameter $J_\tau (\tau)$ versus  $\tau$ for  $\alpha =0.2$ (circle), 0.4(star), 0.6(triangle), 
and $0.8$(square).
 }
\label{super}
\end{figure}
%%%%%%%%%%%%%%%%%%%%%%%%%%%%%%%%%%%%%%%%%%%%%%%%%%%%%%%%%%%%%%%%%%%%%%%%%%%%%%%%%%%%%% 

In order to consider superdiffusion, we used the
above memory enhancement model with  the changed rule as $\sigma_{t} = \sigma_{t-1}$ in the Eq. (\ref{stm})
which induces the superdiffusion with $\gamma = 1+\alpha$ \cite{me}.
We found that $\langle \overline{\delta^2 (\tau)}_\tau \rangle \sim \tau^\gamma$ with $\gamma \approx 1 + \alpha$
as shown in the Fig. \ref{super}, i.e., for superdiffusions $\langle \overline{\delta^2 (\tau)}_\tau \rangle$ 
also provides the 
same anomalous exponents as those obtained by ensemble average. $\langle \overline{\delta^2 (\tau)}_T \rangle$ 
shows the
crossover from the ballistic motion to the normal diffusion indicating the weak ergodicity breaking.
$\langle {\delta^2 (t,\tau)} \rangle$ scales like the subdiffusion and thus the scaling function
of Eq. (\ref{fx}) is also valid for $ \gamma > 1 $ of the superdiffusion (Fig. \ref{super} (c)) . 
Figure \ref{super} (d) shows the EB $J_\tau$ as a function of $\tau$ for $a=10$.
For large $\tau$ they give almost constant value being much smaller than $a=1$ like as in the case of subdiffusion, 
while for small $\tau$ $J_\tau (\tau)$  increases with $\tau$ indicating the small fluctuation resulting from the 
persistence of superdiffusion unlike the subdiffusion
and the larger $\alpha$ is the longer it take to saturate. Thus the time average based on the scaling nature of 
the MSD is also valid for the superdiffusion induced by memory enhancement.

In conclusion, we proposed the solution for the problem caused by the weak ergodicity breaking appeared
in anomalous diffusions by introducing the time-averaged observable based on the scaling law of MSD.
Although the MSD at a aging time depends on the time due to nonstationarity in anomalous
diffusions, scaling nature of the MSD make it be able to obtain the
exponent given by ensemble average through time average for both subdiffusions and superdiffusion.  
While it was found that the heterogeneity of the proposed time-averaged MSD for many realizations depends
on the models and although in the long time limit the heterogeneity is disappeared, in the finite time average 
the heterogeneity may be significant. Thus it need to be considered whether heterogeneity in experiments is due to 
intrinsic nature of a process or just finiteness of interval averaging and it can be easily identified by the parameter $a$
controlling the interval of time average.
This method of averaging over time can be very helpful in elucidating the time scaling underlying in diffusive
phenomena through real experiments and it may be extended to analysis of the other nonstationary processes as well.

\end{document}